\documentclass[12pt]{article}
\begin{document}

\setlength{\baselineskip}{24pt}

\title{
{\Large \bf Fermion Determinant Calculus
}}

\author{
Hisashi {\sc Kikuchi}
\vspace{0.5cm}
\\
{\normalsize\sl Ohu University}\\ 
{\normalsize\sl Koriyama 963, Japan}\\
{\normalsize\tt (kikuchi@yukawa.kyoto-u.ac.jp)}
\vspace{0.2cm}\\
}

\date{}

\maketitle

\begin{abstract} 
The path-integral of the fermionic oscillator with
a time-dependent frequency is analyzed.
We give the exact relation between the boundary condition
to define the domain in which the path-integral is performed and
the transition amplitude that the path-integral calculates.
According to this relation, the amplitude suppressed by a zero mode
does not indicate any special dynamics, 
unlike the analogous situation in field theories. 
It simply says the path-integral picks up
a combination of the amplitudes that vanishes.
The zero mode that is often neglected in the reason of not being 
normalizable is 
necessary to obtain the correct answer for the propagator 
and to avoid an anomaly on the fermion number.
We give a method to obtain the fermionic determinant by the determinant 
of a simple \(2\times 2\) matrix, which enables us
to calculate it for a variety of  boundary conditions.
\end{abstract}

PACS: 11.30.Fs; 03.65.-w; 02.70.Hm; 11.30.Pb
\vspace{0.5cm}

\newcommand{\D}{{\cal D}}
\newcommand{\be}{\begin{equation}}
\newcommand{\ee}{\end{equation}}

Fermionic determinant of the operator \(\D\) is the  one we first encounter 
in the analysis of the quantum physics in the path-integral formalism.
It is the exact result  of the Grassmann path-integral made from
a bilinear  Lagrangian  \(\bar\psi \D \psi\) over the fermionic degrees of freedom
 \(\psi\) and \(\bar\psi\).
The determinant carries an important information about the time evolution
of the fermions under the influence of bosonic background.
Especially  when  \(\D\) or its adjoint \(\D^\dagger\)   has a zero mode,  
the zero-frequency eigenmode of the operator,  
the determinant  vanishes and corresponding
transition is suppressed.
A typical example of such situation happens with 
\(\D\) the Dirac operator in SU(2) gauge theory.
It possesses zero modes in the instanton background.
The consequent suppression of the transition  is
interpreted  to reflect the fermion number violation
due to the anomaly on the fermion current \cite{'tHooft}.

The physical significance of the zero modes of Dirac operators was
first advocated in Ref.~\cite{JR} and has been discussed in various places
in physics. 
We still notice there seems to be a confusion on the 
treatment of zero modes in  the case of
fermionic oscillator, the simplest system containing only one
fermionic degree of freedom.
Its  Lagrangian \(\bar\psi \D \psi\) in imaginary time formalism 
is given by a simple first order differential operator
\be  \D  =  {d\over d\tau} + v(\tau)
\ee
with respect to the imaginary time \(\tau\),
where \(v(\tau)\) is the time-dependent angular frequency induced by
a coupling to a bosonic degree of freedom.
We also define
\be \D^\dagger  = - {d\over d\tau} + v(\tau).\ee
Assume  \(v(\tau)\) changes its sign along the
evolution in the time interval \([-T/2, T/2]\), for example, as 
\(v(\tau) = \tanh( \tau ) \).
The solution of \(\D \varphi = 0 \) is then regarded as the zero mode
since it is normalizable in the limit of \(T \rightarrow \infty\).
Gildener and Patrascioiu have argued by an explicit calculation
that there is no zero mode available in the determinant calculus
even in this simple example \cite{Gildener},  while 
Salomonson and Van Holten have taken advantage of 
the zero mode in their calculation for the 
supersymmetry breaking \cite{Salomonson}.

In addition to the question of the existence of zero mode,
we are concerned that the fermionic oscillator would have an anomaly
 if the numbers of the zero modes of 
\(\D\) and \(\D^\dagger\) are different.
In the simple example mentioned above, the solution
of \(\D^\dagger\phi=0\) is not thought to be the relevant zero mode:
for sufficiently large \(T\), it becomes zero almost everywhere 
when normalized to one.
Thus there appears to be the asymmetry in the numbers.
According to the path-integral formulation of the anomaly \cite{Fujikawa},  
this asymmetry induces a phase in the path-integral measure
under the global phase transformation,
\(\psi \rightarrow e^{i\theta} \psi\) and 
\(\bar\psi \rightarrow e^{-i\theta} \bar\psi\),
which indicates the non-conservation of the fermion number.
This  contradicts our naive intuition,
that the fermionic oscillator should have
no anomaly since we can  calculate
any amplitude with no regularization.

Motivated by this question of zero mode, we will clarify in
this letter the exact calculation of the path-integral
of the fermionic oscillator.
In doing the path-integral, we first prepare an orthonormal 
complete set for all possible configurations.
The eigenmodes of a certain self-adjoint operator are known 
to constitute such a complete set.
The useful choice for the calculation is to use
the differential expressions \(\D^\dagger \D\) and 
\(\D \D^\dagger\) as the candidate for the operator.
As is known in mathematics, we need to impose the boundary
condition to make the differential expressions to be self-adjoint.
The important observation found in this letter is
that the path-integral represents a different transition
amplitude if the boundary condition is different.
We will give the exact relation between the boundary condition and
the transition amplitude, and verify it by comparing the path-integral 
results with those obtained by the operator formalism.
The suppression for the zero mode will be understood 
within this boundary condition dependence.

\newcommand{\zeroket}{|0\rangle}
\newcommand{\zerobra}{\langle 0 |}
\newcommand{\oneket}{|1\rangle}
\newcommand{\onebra}{\langle 1 |}

We first carry out the calculation in the operator formalism.
The Hamiltonian of the oscillator is
\be
H(\tau) = v(\tau) {1\over 2} \left(\Psi^\dagger \Psi - 
\Psi \Psi^\dagger \right), \label{H}
\ee
where \(\Psi\) and \(\Psi^\dagger\) are the annihilation
and creation operators  in the two-dimensional space spanned by
vacant \(\zeroket\) and occupied \( \oneket \) states,
\be
\Psi \zeroket = \Psi^\dagger \oneket = 0,\quad
\Psi \oneket = \zeroket,   \quad \Psi^\dagger \zeroket = \oneket.
\label{DPsi}
\ee
From this Hamiltonian, the evolution operator from the initial time
 \( T_i\) to the final \(T_f\) is obtained  by
\be U(T_f, T_i) = {\cal T}\exp
\left[-\int_{T_i}^{T_f}d\tau' H(\tau') \right],
\ee
where \(\cal T\) represents the time-ordered product.
The Hamiltonian commutes with the fermion number
operator \(\Psi^\dagger \Psi\),  and it is obvious
that the matrix elements of \(U\) are written as
\be
\zerobra U(T_f, T_i) \zeroket  =  \exp\left[  {1\over 2} 
\int^{T_f}_{T_i} d\tau \, v(\tau) \right], \quad
\onebra U(T_f, T_i) \oneket = \exp\left[  - {1\over 2} 
\int^{T_f}_{T_i} d\tau \,v(\tau) \right], \label{ex2}
\ee
and two off-diagonal ones equal to zero.
These are all we have to know to obtain any transition amplitude.

We turn to the path-integral formalism and consider 
\be
I  =  \int [d\bar\psi d\psi ] \exp
\left[-\int_{-T/2}^{T/2} d\tau \bar\psi \D \psi \right],\label{I}
\ee
where the functional measure is defined by  
discretizing imaginary-time \cite{Faddeev}.
We will  come back this point later.
Eq. (\ref{I}) is related to the matrix elements of \(U(T/2, -T/2)\).
Before seeing it, let us proceed the calculation of Eq.~(\ref{I}) usually 
adopted in literatures.
The most decent way of the calculation for non-Hermitian
\(\D\) such as the one in the present case is to choose the domains 
\(D_\varphi\) and \(D_\phi\) of the square-integrable functions
in the interval \([-T/2,T/2]\), in which
\(\D^\dagger  \D\) and \(\D \D^\dagger\) are  self-adjoint and positive semi-definite,
and the non-zero eigenmodes have the one-to-one correspondence 
(see for example \cite{Fujikawa}).
The normalized eigenmodes \(\varphi^{(n)} (\in D_\varphi ) \)  and 
\(\phi^{(n)} (\in D_\phi) \) (\(n = 1, 2, 3, ....) \),
\be 
\D^\dagger \D \varphi^{(n)} = \lambda_n \varphi^{(n)},\quad 
\D \D^\dagger \phi^{(n)} = \lambda_n \phi^{(n)},\label{defdiff}
\ee
constitute a complete orthonormal set and they are related
by
\be {1 \over \sqrt \lambda_n } \D \varphi^{(n)} = \phi^{(n)}, \quad
{1 \over \sqrt \lambda_n } \D^\dagger \phi^{(n)} = \varphi^{(n)},
\label{1T1}\ee
except zero modes.
The eigenvalues so obtained are non negative.
Using the expansions with these eigenmodes,
\be
\psi(\tau) = \sum_n a_n \varphi^{(n)}(\tau), \quad 
\bar\psi(\tau) = \sum_n \bar a_n \phi^{(n)*}(\tau) ,\label{anan}\ee
we employ the integration measure \([d\bar a_n\,d a_n]\) instead of 
\([ d\bar \psi\, d\psi]\) 
 and  obtain
\be 
I = {\cal N} \int [d\bar a d a ] 
e^{ - \sum_n \sqrt \lambda_n \bar a_n a_n }
 = {\cal N} \left[ \det(\D^\dagger \D) \right]^{1/2}, \label{Irst}
\ee
where  \(\cal N\) is the Jacobian between the measures,
\(\det(\D^\dagger \D) \) is the infinite product of the eigenvalues.
The latter is  well-defined  in the combination with \(\cal N\).
A different background \(v(\tau)\) results in a
different complete set and different amplitudes in Eq.~(\ref{anan}),
say \(b_n\) and \(\bar b_n\).
The Jacobian between the measures \( [d\bar a da ]\) and \([d\bar b db]\)
is the determinant of the matrix that represents the linear transformation between
the eigenmodes in the different sets.
Since each set is  complete and orthonormal  anyway,  the
matrix is unitary and the Jacobian is a phase.
Thus background dependence of \(\cal N\) in (\ref{Irst}) can be at most a phase
 factor\footnote{%%
The procedure described here has a greater advantage in this property 
than the direct use of 
the eigenmodes of non self-adjoint \(\D\)  as was done in Ref.~\cite{Gildener}.}.
In the explicit examples adopted in later calculations, however,
we can choose the eigenmodes as real function of \(\tau\).
This is because either the differential equations (\ref{defdiff}) or
boundary conditions (see  Eqs.~(\ref{Bvarphi}) and (\ref{Bphi})) 
to solve the eigenvalue problem does not have any imaginary variable or constant.
Thus the phase factor is \(1\) or \(-1\).
The Jacobian \(\cal N\) then becomes background-independent
except a possibility that it changes the sign at some special background.

Although the differential expression 
\(\D^\dagger \D\) or \(\D \D^\dagger\) seems self-adjoint by its form,
it is  not immediately so.
One needs to define the  domains \(D_\varphi\) and \(D_\phi\) properly in
the functional space by  imposing the boundary condition.
We notice that any \(\varphi_1\) and \(\varphi_2\) in \(D_\varphi\) should obey
\be \varphi_2^*(-T/2) \D \varphi_1(-T/2) - \varphi_2^*(T/2) D \varphi_1(T/2) = 0.
\label{BC} \ee
Under the usual definition of the inner product,
\be (\varphi_2, \varphi_1) \equiv \int_{-T/2}^{T/2} d\tau \varphi^*_2(\tau)\varphi_1(\tau),
\ee
Eq.~(\ref{BC}) means 
\(( \varphi_2, \D^\dagger \D \varphi_1 ) = (\D \varphi_2, \D\varphi_1)\)
which guarantees that \(\D^\dagger\D\) is non-negative, and leads to
\( (\varphi_2, \D^\dagger \D \varphi_1) = (\D^\dagger\D \varphi_2, \varphi_1)
\)
which  holds if it is self-adjoint.
Note also that the equation
\be \int^{T/2}_{-T/2} d\tau \bar \psi \D\psi = \int^{T/2}_{-T/2}d\tau (\D^\dagger\bar\psi) \psi.
\ee
holds as far as we  expand \(\psi\) and \(\bar\psi\) as in (\ref{anan}) with
\(\varphi\) and \(\phi\) that obey (\ref{1T1}).
This equation means that any \(\varphi \in D_\varphi\) and any \(\phi \in D_\phi\)
satisfy 
\be \phi^*(-T/2) \varphi(-T/2) - \phi^*(T/2) \varphi(T/2) = 0.\label{ABC}\ee
Since \(\D\varphi \in D_\phi\) if \(\varphi \in D_\varphi\),
Eq.~(\ref{BC}) and (\ref{ABC}) are equivalent.

In fact the mathematical theory of the self-adjoint extension of the differential expression 
tells us that we need to specify  two  linearly independent boundary conditions on the
values, \(\varphi(-T/2)\) and \(\varphi(T/2)\), and the first derivatives, \(\dot\varphi(-T/2)\)
and \(\dot\varphi(T/2)\) in the present case of the second order differential expression.
We refer the readers to the mathematical textbook \cite{Akhiezer} about the
details.
The problem is that the boundary condition so obtained is not unique.
Then there naturally occurs a question about the boundary condition
dependence of the path-integral.

To see the boundary condition dependence,
we review the derivation of Eq.~(\ref{I}) based on Ref.~\cite{Faddeev}.
We prepare the states
\be |\theta\rangle \equiv \zeroket +  \oneket \theta, \quad 
\langle \bar\theta| \equiv \zerobra + \bar\theta \onebra
\ee
making use of Grassmann numbers  \(\theta\) and \(\bar\theta\).
They satisfy the completeness relation,
\be \int d\bar\theta d \theta (1-\bar\theta \theta)|\theta\rangle
\langle \bar\theta| = \zeroket \zerobra  + \oneket \onebra. \label{complete}
\ee
The path-integral (\ref{I}) calculates the evolution from
\( | \theta_0 \rangle\) to \( | \bar\theta_N \rangle\),
\(\langle \bar \theta_N | U| \theta_0 \rangle\),
where we have abbreviated \(U(T/2, -T/2) \) to \(U\).
To evaluate this, we discretize the time interval into \(N\) segments,
each of which has the length \(\epsilon = T/N\) and write \(U\) as
\be U = \lim_{N \rightarrow \infty} {\cal T} 
\left( \prod_{n=1}^N (1 - \epsilon
H(\tau_n) )\right), \ee
where \(\tau_n = n\epsilon - T/2 \).
We insert \((N-1)\) pairs of Grassmann integrals of
\((\bar \theta_n, \theta_n) (n = 1, ... , N-1) \)
in the form of the completeness relation (\ref{complete}) as 
the \((N-1)\) junctions of the \(N\) factors \((1 - \epsilon H(\tau_n))
(n = 1, ... , N)\).
Using the explicit form (\ref{H}) and (\ref{DPsi}), we obtain
\begin{eqnarray}
\langle \bar \theta_N | U| \theta_0 \rangle &  = & \lim_{N \rightarrow \infty}
 \int d\bar\theta_{N-1} d\theta_{N-1} ... d\bar\theta_1 d\theta_1 \nonumber\\
&&\times \left[
1 + \epsilon {v(\tau_N) \over 2}+\left(1 - \epsilon {v(\tau_N) \over 2}\right) 
\bar\theta_N\theta_{N-1}
\right] \nonumber\\
&&\times\left[
1 + \epsilon {v(\tau_{N-1}) \over 2} - \left(1 + \epsilon {v(\tau_{N-1}) \over 2}\right)
\bar\theta_{N-1} \theta_{N-1}
+ \left(1 - \epsilon {v(\tau_{N-1}) \over 2}\right) \bar\theta_{N-1}\theta_{N-2}
\right] \nonumber\\
&&\times ...\nonumber\\
&& \times\left[
1 + \epsilon {v(\tau_1) \over 2} - \left(1 + \epsilon {v(\tau_1) \over 2}\right)
\bar\theta_1 \theta_1
+ \left(1 - \epsilon {v(\tau_1) \over 2}\right) \bar\theta_1\theta_0
\right]. \label{derivation}
\end{eqnarray}
All the bracket factors except the first one have the form
\be \left[
1 + \epsilon {v(\tau_i) \over 2} - \left(1 + \epsilon {v(\tau_i) \over 2}\right)
\bar\theta_i \theta_i
+ \left(1 - \epsilon {v(\tau_i) \over 2}\right) \bar\theta_i\theta_{i-1}
\right], \quad (i = 1, ..., N-1) \ee
and can be safely replaced with
\be \exp\left\{
 - \epsilon \bar\theta_i \left[ \left({\theta_i -\theta_{i-1}\over \epsilon}
\right) + v(\tau_i)
\left({\theta_i + \theta_{i-1}\over 2}\right)
\right]\right\}
\ee
because \(\bar\theta_i^2 = 0\) and \(\int d\bar\theta_i = 0\).
Then we arrive at a very close  expression to Eq.~(\ref{I}) when
we regard \(\theta_n\) as \(\psi(\tau_n)\) and \(\bar\theta_n\) as
\(\bar\psi(\tau_n)\).
Note, however, a difference of the first bracket
factor in Eq.~(\ref{derivation}) from the others.
We also wonder where \(\bar\theta_N\) and \(\theta_0\) have gone
when we got the result (\ref{Irst}).

These questions are solved naturally by assuming that 
the exact definition of the measure \([d\bar\psi d\psi]\) 
in Eq.~(\ref{I}), corresponding to the way we have integrated it out, does 
include the integral over \(\bar\theta_N\) and \(\theta_0\) as
\be I \equiv \int d\bar\theta_N d\theta_N ( 1- \bar\theta_N \theta_N )
\langle \bar\theta_N| U|\theta_0\rangle \label{Idef}\ee
 where \(\theta_0\) and \(\theta_N\) 
are related by the boundary condition to make \( \D^\dagger \D \) and
\(\D \D^\dagger\) self-adjoint.
The relevant terms in the integrand in (\ref{Idef}) 
is those quadratic in the Grassmann variables,
\be
(1-\bar\theta_N\theta_N)\langle \bar\theta_N| U|\theta_0\rangle
=... - \bar\theta_N\theta_N \zerobra U \zeroket + 
\bar\theta_N\theta_0 \onebra U \oneket + ...\label{intd}
\ee
This shows 
\be
 I =  \zerobra U \zeroket + \beta \onebra U \oneket
\label{ME}\ee
if  the boundary conditions imposed to define \(D_\varphi\) is
\( \theta_0 + \beta \theta_N = 0 \), or equivalently
\be \varphi(-T/2) + \beta \varphi(T/2) = 0. \label{BB}\ee
Eqs.~(\ref{ME}) and (\ref{BB}) are consistent with the fact
that the fermionic path-integral presents the trace of \(U\) when
carried out in the anti-periodic configurations (\(\beta = 1\)).

We verify Eq.~(\ref{ME})  by explicitly calculating the determinant \(\det (\D^\dagger \D)\)
at various boundary conditions.
The boundary conditions we are interested in is written
as 
\be  \varphi(-T/2) + \beta \varphi(T/2) = 0, \quad \beta \D\varphi(-T/2) + \D\varphi(T/2)
=0,  \label{Bvarphi}\ee
and 
\be  \beta \phi(-T/2) + \phi(T/2) = 0, \quad \D^\dagger \phi(-T/2) + \beta \D^\dagger
\phi(T/2)=0, \label{Bphi}\ee
with a real parameter \(\beta\). 
This condition includes the periodic one at \(\beta = -1\) as well as the
anti-periodic at \(1\). It is also available for the domain in which the zero mode lives.
Once we put the boundary condition on \(\varphi\), Eq.~(\ref{BC}) and the one-one
correspondence between \(\varphi\) and \(\phi\)
determine the conditions on \(\D\varphi\), \(\phi\), and \(\D^\dagger \phi\).
The operators \( \D^\dagger \D\) and \( \D \D^\dagger\)
are proved to be self-adjoint under these boundary conditions \cite{Akhiezer}.
To calculate \( \det(\D^\dagger \D)\),
the \(2\times 2\) matrix
\be M (z) \equiv \left(\begin{array}{cc}
u_1(z; -T/2) + \beta u_1(z; T/2) & u_2(z; -T/2) + \beta u_2(z; T/2) \\
\beta \D u_1(z; -T/2) + \D u_1(z; T/2) & \beta \D u_2(z; -T/2) + \D u_2(z; T/2)
\end{array}
\right)\label{Mmp}\ee
plays the central role,
where \(u_i(z; \tau)\, ( i = 1, 2 ) \) are the linearly independent solutions of the 
equation
\be 
\D^\dagger \D u_i(z; \tau) = z u_i(z;\tau) \label{Eqforu}
\ee
and the parameter \(z\) is  complex in general.
We fix the normalization of these solutions by
\begin{eqnarray}
 u_1(z; -T/2) = 1 &\quad & u_2(z;-T/2) = 0 \nonumber \\
\dot u_1(z;-T/2) = 0 &\quad &\dot  u_2(z;-T/2) = 1 .\label{Bforu}
\end{eqnarray}
Note the differential equation (\ref{Eqforu}) does not have
the first derivative term. Thus the Wronskian (\(u_1 \dot u_2 
- \dot u_1 u_2 \)) conserves and is equal to \(1\) at any \(\tau\).
The zero points of \(\det M\) coincide with the eigenvalues 
of \(\D^\dagger \D\):
if  \(\lambda\) is one of the  eigenvalues,
there exists a non-trivial linear combination
\(\gamma_1 u_1(\lambda; \tau) + \gamma_2 u_2(\lambda;\tau)\)
that satisfies Eq.~(\ref{Bvarphi});
the equation for \(\gamma_i \) turns out to be \newline \( M(\lambda)_{ij}\, \gamma_j = 
0\, (i,j = 1,2) \)
and thus \(\det M(z)\) is zero at the eigenvalue \(\lambda\);
one can reverse this argument in the opposite direction.

We then see the ratio \([\det(\D^\dagger \D -z )/ \det M(z)]\)
is independent from the background \(v(\tau)\).
The proof is essentially the same  as the one given in  Ref.~\cite{Coleman}
in the calculation of the other type of  determinants.
Let us consider two different operators \(\D_1^\dagger \D_1\) and 
\(\D_2^\dagger \D_2\) containing different backgrounds
\(v_1(\tau)\) and \(v_2(\tau)\), and denote their \(n\)-th eigenvalue
by  \(\lambda_{1;n}\) and \( \lambda_{2;n}\), respectively.
Correspondingly let \(M_{1}\) and \(M_{2}\) denote the
matrix made by (\ref{Mmp}) and (\ref{Eqforu})
with \(\D_1^\dagger \D_1\) and  \(\D_2^\dagger \D_2\), respectively.
The ratio defined by 
\be {\det(\D_1^\dagger \D_1 -z )  \over \det(\D_2^\dagger \D_2 -z )}
\equiv \prod _{n = 1}^\infty 
\left({\lambda_{1;n}-z \over \lambda_{2;n}-z }\right) \label{ratio}\ee
is a meromorphic function of \(z\), and it
has a simple zero at each \(\lambda_{1;n}\) and a simple pole at
each \(\lambda_{2;n}\). 
It goes to one as \(z\) goes to infinity in any direction except along
the real positive axis.
The ratio \([\det M_1(z) / \det M_2(z)]\) is also a meromorphic function that
has poles and zeros at exactly 
the same \(z\).
Note further we obtain
\be \det M(z) = \D u_2(z;T/2) + \beta^2 \left[ u_1(z; T/2) - v(-T/2) u_2(z; T/2)\right]
+ 2 \beta \label{detM}\ee
using the condition (\ref{Bforu}) and the conservation of the Wronskian.
For sufficiently large \(|z|\), \(\sqrt{|z|} \gg |v(\tau)^2 - \dot v(\tau)| T \),
the frequency \(v(\tau)\) in (\ref{Eqforu})  becomes negligible.
The solutions \(u_i(z;\tau)\) is then well-approximated by their free solutions,
and \(\dot u_2(T/2) \simeq u_1(T/2) \simeq e^{\sqrt{-z} T }/2 \).
The determinant of \( M\) then grows as \( (1+\beta^2) e^{\sqrt{-z} T }/2\) at sufficiently
large \(|z|\) (except along the real positive axis)
independently from the backgrounds.
The ratio \([\det M_1(z) / \det M_2(z)]\) also goes to one in the same limit.
Thus
\be {\det(\D_1^\dagger \D_1 -z )  \over \det(\D_2^\dagger \D_2 -z )}
= {\det M_1(z) \over \det M_2(z)}. \label{bind}\ee
Eq.~(\ref{bind}) establishes  that \( [\det(\D^\dagger \D -z )/ \det M(z)]\) is
background independent.

We can now write\footnote{%
Similar formulae that relate the determinant of differential
operators with that of a matrix have been found in 
condensed matter physics \cite{Nakahara}.}
\be {\cal N} [\det(\D^\dagger \D )]^{1/2}  = {\cal N'} \left[ \det M(0)\right] ^{1/2}.
\label{detDdetM}\ee
The factor \(\cal N'\) is  background independent as 
is \(\cal N\).
The calculation of \(\det M(0)\) is elementary.
We obtain
\be
u_1(0;\tau) = x_1(\tau) + v(-T/2) x_2(\tau) , \quad
u_2(0;\tau) = x_2(\tau), \label{us}
\ee
where 
\be
x_1(\tau)  =   \exp\left[ - \int_{-T/2}^\tau d\tau' v(\tau') \right], \quad
y_1(\tau)  =   \exp\left[  \int_{-T/2}^\tau d\tau' v(\tau') \right], \quad
x_2(\tau)  =   x_1(\tau) \int_{-T/2}^\tau d\tau' \left(y_1(\tau')\right)^2.
\label{xyz}
\ee
Putting these solutions into (\ref{detM}), we  find
\be \det M(0) = 
\left( [y_1(T/2) ] ^{1/2} + \beta [x_1(T/2)]^{1/2}\right)^2. \label{detM0}
\ee
Note \(y_1 = (x_1)^{-1}.\)
Recalling Eqs.~(\ref{Irst}), (\ref{detDdetM}), and (\ref{detM0}), we obtain
\be I  = {\cal N'} \left\{ [y_1(T/2) ] ^{1/2} + \beta [x_1(T/2)]^{1/2} \right\}. \label{Ipath}
\ee
This is exactly  Eq.~(\ref{ME}), where the matrix elements are calculated
explicitly by Eqs.~(\ref{ex2}) and (\ref{xyz}).
The factor \(\cal N'\) turns out to be \(1\).
We like to make a comment here. \(I\) in Eq.~(\ref{Ipath}) becomes negative, for example,
when \(\beta = -1\) and \(x_1(T/2) > y_1(T/2) \).
It means that we have chosen the negative solution  in taking the square-root of \(\det M\)
in this case. This choice is justified by considering that \(I\) should be
analytic with respect to the functional variation of \(v(\tau)\).
The flow of \(\sqrt \lambda_n\) that appears in (\ref{1T1}) and (\ref{Irst})
must be smooth when the background changes continuously.

We are prepared to answer the  question of zero mode.
Note that the zero mode candidate is \(x_1\) or \(y_1\) in (\ref{xyz}).
Let us start with assuming that the normalizable one in the
usual sense is \(x_1\). 
The domain in which \(x_1\) resides is given by the boundary condition (\ref{Bvarphi})
with \(\beta = - y_1(T/2)\).
Now that zero is an eigenvalue, the path-integral is zero.
This is readily verified by putting the value of \(\beta\) into (\ref{Ipath}).
The reason of the vanishing amplitude is, however, quite different from
our interpretation of the same occurrence in the SU(2) gauge theory.
There is no dynamical reason;
we have just chosen a vanishing combination accidentally
by the boundary conditions.

We also notice the other zero solution \(y_1\) cannot be
neglected.
It satisfies the boundary condition  Eq.~(\ref{Bphi}) with
the same value of \(\beta\).
Since we can normalize it any way as long as the time interval
\(T\) is finite (no matter how long it is), we cannot find any legitimate reason to
abandon it. 
We can confirm the necessity of \(y_1\) in the calculation of 
the propagator.
Let us define
\begin{eqnarray}
 F(\tau, \tau') &\equiv &\theta(\tau - \tau') U(T/2, \tau) \Psi U(\tau, \tau') \Psi^\dagger
U(\tau', -T/2) \nonumber\\
&&- \theta(\tau'-\tau) U(T/2, \tau')\Psi^\dagger U(\tau',\tau) 
\Psi U(\tau, -T/2),
\end{eqnarray}
and consider its path-integral representation 
\be G(\tau, \tau') = \int [d\bar\psi d\psi] \exp\left[-\int_{-T/2}^{T/2} d\tau \bar\psi\D\psi \right] \, 
\psi(\tau) \bar\psi(\tau') \label{PIfG}.\ee
The exact relation of \(G\) integrated in the domain defined by Eqs.~(\ref{Bvarphi}) 
and (\ref{Bphi}) 
to the corresponding matrix element of \(F\)
is obtained by applying the same argument that leads to Eq.~(\ref{ME}). It is 
\begin{eqnarray}
 G(\tau, \tau') & = & \theta(\tau - \tau')\,
\zerobra U(T/2, \tau) \Psi U(\tau, \tau') 
\Psi^\dagger U(\tau', -T/2) \zeroket  \nonumber\\
&&+ \theta(\tau'-\tau)\, y_1(T/2) \onebra U(T/2, \tau')\Psi^\dagger U(\tau',\tau) \Psi
U(\tau,-T/2) \oneket.\label{G}
\end{eqnarray}
In the operator formalism, Eqs.~(\ref{DPsi}) and (\ref{ex2}) yield
\begin{eqnarray}
 \zerobra U(T/2, \tau) \Psi U(\tau, \tau') \Psi^\dagger U(\tau', -T/2) \zeroket &= &
y_1(T/2) ^{1/2} x_1(\tau) y_1(\tau'), \nonumber\\
\onebra U(T/2, \tau) \Psi^\dagger  U(\tau', \tau) 
\Psi U(\tau, -T/2) \oneket & = & x_1(T/2) ^{1/2} x_1(\tau) y_1(\tau'),\label{PP}
\end{eqnarray}
and, thus, using Eq.~(\ref{G}) we obtain
\be G(\tau, \tau') = y_1(T/2)^{1/2}x_1(\tau) y_1(\tau') \label{G0xy}.\ee
Interestingly the final result does not have any remnant of 
the time-ordered procedure in (\ref{G}).

We do the corresponding  path-integral (\ref{PIfG}).
The normalized zero modes are
\be
 \varphi^{(1)}(\tau)  =  {x_1(\tau) \over 
{\displaystyle \sqrt{\int^{T/2}_{-T/2} d\tau'\, x_1(\tau')^2} }}, \quad
\phi^{(1)}(\tau)  =  {y_1(\tau) \over 
{\displaystyle \sqrt{\int^{T/2}_{-T/2} d\tau' \, y_1(\tau')^2} }}. 
\label{zeromode}
\ee
The Grassmann variables \(a_1\) and \(\bar a_1\) 
in the expansion (\ref{anan}),
the coefficients of \(\varphi^{(1)}\) and \(\phi^{(1)}\),
do not appear in the action.
Only the \(\psi\) and \(\bar \psi\) in the integrand in (\ref{PIfG}) can supply 
them, and one gets
\be
 G(\tau,\tau') = {\cal N} \left[ \det{}'(\D^\dagger \D)\right] ^{1/2} 
\varphi^{(1)}(\tau)
\phi^{(1)}(\tau'),\label{PIresG}\ee
where \(\det{}' (\D^\dagger \D)\) is the product of the eigenvalues except zero.
It is evaluated by
\be {\cal N}^2 \det{}'(\D^\dagger \D) = \lim_{z\rightarrow 0} 
{\det M(z) \over (-z)}. \label{det'}
\ee
For the calculation of the right-hand side in Eq.~(\ref{det'}),
we use the fact that the solutions \(u_i(z;\tau) (i=1,2)\) have an expansion 
around \(z=0\), 
\be u_i(z;\tau) = u_i(0;\tau) + z\, \delta u_i(\tau) + ...
\label{exofu}\ee
and the first order term is given by
\be \delta u_i(\tau) = \int ^\tau_{-T/2} d\tau' \left[ u_1(0;\tau) u_2(0;\tau') 
- u_2(0;\tau) u_1(0;\tau')\right]
 u_i(0;\tau'). \label{delu}\ee
Since \(\delta u_i(-T/2) = \delta \dot u_i(-T/2) = 0\), 
the expansion (\ref{exofu}) satisfies  the initial conditions 
(\ref{Bforu}).
Putting Eqs.~(\ref{exofu}) and (\ref{delu}) into Eq.~(\ref{detM}) and using
Eqs.~(\ref{us}) and (\ref{xyz}), we obtain
\be \lim_{z\rightarrow 0} {\det M(z) \over (-z) }
 = y_1(T/2)\, \left[\int^{T/2}_{-T/2} d\tau\, y_1(\tau)^2 \right] 
\left[ \int^{T/2}_{-T/2} d\tau\, x_1(\tau)^2 \right] 
\label{detp}.\ee
Eqs.~(\ref{zeromode}), (\ref{PIresG}), (\ref{det'}) and (\ref{detp}) give
the correct result (\ref{G0xy}).
We would have a wrong answer without \(\phi^{(1)}\).
The numbers of the zero mode belonging to 
\(D_\varphi\) and \(D_\phi\) are the same independently
of the specific boundary condition to define them.
This is consistent with the absence of the anomaly in the 
fermionic oscillator.

In summary,
we have revealed and confirmed that the path-integral in different boundary conditions 
calculates different matrix elements of the time evolution operator.
The exact relation between the boundary condition and the matrix element
tells us the reason of vanishing path-integral by the zero mode.
It does not have a dynamical reason such as the fermion number violation,
but it occurs just because the zero mode takes the vanishing combination
of the matrix elements.
One may wonder whether the boundary condition dependence  vanishes
in the limit of \(T\rightarrow \infty\).
It does not vanish.
For example, the path-integral with the periodic boundary condition 
can be either positive or negative depending on the background,
while that with the anti-periodic one is always positive no
matter what the background is.
It is natural to expect the existence of the similar boundary condition 
dependence in field theories.
Then any conclusion based on a path-integral calculation should be spelled out with
the boundary condition, with respect to the imaginary time, explicitly specified.
The conclusion might change for different boundary conditions.
As far as we know there is little discussion on this dependence.
A careful investigation of field theories taking consciously
boundary conditions into account may reveal their new aspects.

\begin{center}
{\bf Acknowledgment}
\end{center}
The author thanks  M.~Nakahara, H.~Aoki, and H.~Aoyama for the valuable discussions.
M.~Nakahara also let him know about the related works 
on the determinant calculation.
He also thanks to the hospitality at KEK that 
enabled him to stay there for this work 
as a visiting research scientist.

\newcommand{\J}[4]{{\sl #1} {\bf #2} (19#3) #4}
\newcommand{\MPL}{Mod.~Phys.~Lett.}
\newcommand{\NP}{Nucl.~Phys.}
\newcommand{\PL}{Phys.~Lett.}
\newcommand{\PR}{Phys.~Rev.}
\newcommand{\PRL}{Phys.~Rev.~Lett.}
\newcommand{\AP}{Ann.~Phys.}
\newcommand{\CMP}{Commun.~Math.~Phys.}
\newcommand{\CQG}{Class.~Quant.~Grav.}
\newcommand{\PRP}{Phys.~Rept.}
\newcommand{\SPU}{Sov.~Phys.~Usp.}
\newcommand{\RMPA}{Rev.~Math.~Pur.~et~Appl.}
\newcommand{\SPJ}{Sov.~Phys.~JETP}
\newcommand{\MP}{Int.~Mod.~Phys.}
\newcommand{\JMP}{J.~Math.~Phys.}
\newcommand{\PTP}{Prog.~Theor.~Phys.}

\end{document}